\documentclass[sigplan,10pt]{acmart}

\makeatletter
\newcommand{\printfnsymbol}[1]{%
  \textsuperscript{\@fnsymbol{#1}}%
}
\makeatother

\usepackage{tikz}
\usepackage{amsmath}
\usepackage{outlines}
\usepackage[colorinlistoftodos]{todonotes}
\usepackage{subcaption}
\usepackage{filecontents}
\usepackage{comment}
\usepackage{xspace}
\usepackage{mwe}
\usepackage{booktabs}
\usepackage{graphicx}
\usepackage{multirow}
\usepackage{makecell}

\usepackage[utf8]{inputenc}

\usepackage[shortcuts]{extdash} %

\usepackage{breakurl} %

\usepackage{enumitem}
\setlist{itemsep=1pt,topsep=2pt,partopsep=0pt,parsep=1pt,listparindent=0pt,leftmargin=1.5pc}

\usepackage{etoolbox} %

\setcopyright{none}
\renewcommand\footnotetextcopyrightpermission[1]{} %
\usepackage{listings}
\colorlet{punct}{red!60!black}
\definecolor{background}{HTML}{EEEEEE}
\definecolor{delim}{RGB}{20,105,176}
\colorlet{numb}{magenta!60!black}

\lstdefinelanguage{json}{
basicstyle=\scriptsize\ttfamily,
numbers=left,
numberstyle=\scriptsize,
stepnumber=1,
numbersep=8pt,
showstringspaces=false,
breaklines=true,
frame=lines,
backgroundcolor=\color{background},
literate=
*{0}{{{\color{numb}0}}}{1}
{1}{{{\color{numb}1}}}{1}
{2}{{{\color{numb}2}}}{1}
{3}{{{\color{numb}3}}}{1}
{4}{{{\color{numb}4}}}{1}
{5}{{{\color{numb}5}}}{1}
{6}{{{\color{numb}6}}}{1}
{7}{{{\color{numb}7}}}{1}
{8}{{{\color{numb}8}}}{1}
{9}{{{\color{numb}9}}}{1}
{:}{{{\color{punct}{:}}}}{1}
{,}{{{\color{punct}{,}}}}{1}
{\{}{{{\color{delim}{\{}}}}{1}
{\}}{{{\color{delim}{\}}}}}{1}
{[}{{{\color{delim}{[}}}}{1}
{]}{{{\color{delim}{]}}}}{1},
}

\usepackage{algorithm}
\usepackage[noend]{algpseudocode}
\makeatletter
\renewcommand{\ALG@beginalgorithmic}{\footnotesize}
\algnewcommand{\LeftComment}[1]{\Statex \(\triangleright\) #1}
\makeatother

\definecolor{codegreen}{rgb}{0,0.6,0}
\definecolor{codeblue}{rgb}{0,0.5,1.0}
\definecolor{codegray}{rgb}{0.5,0.5,0.5}
\definecolor{codepurple}{rgb}{0.58,0,0.82}
\definecolor{backcolour}{rgb}{0.95,0.95,0.92}

\newcommand{\myparagraph}[1]{\noindent\textbf{#1}}
\newcommand{\SystemName}{\textsc{Llama}\xspace}

\begin{document}
\title{\textsc{\SystemName:} A Heterogeneous \& Serverless Framework for Auto-Tuning Video Analytics Pipelines}

\author{Francisco Romero}
\affiliation{%
  \institution{Stanford University}
  \country{}
}
\authornote{Equal contribution}
\author{Mark Zhao\printfnsymbol{1}}
\affiliation{%
  \institution{Stanford University}
  \country{}
}
\author{Neeraja J. Yadwadkar}
\affiliation{%
  \institution{Stanford University}
  \country{}
}
\author{Christos Kozyrakis}
\affiliation{%
  \institution{Stanford University}
  \country{}
}

\date{}

\begin{abstract}
The proliferation of camera-enabled devices and large video repositories has led to a diverse set of video analytics applications.
These applications rely on video pipelines, represented as DAGs of operations, to transform videos, process extracted metadata, and answer questions like, ``Is this intersection congested?''
The latency and resource efficiency of pipelines can be optimized using configurable knobs for each operation (e.g., sampling rate, batch size, or type of hardware used).
However, determining efficient configurations is challenging because
(a) the configuration search space is exponentially large, and
(b) the optimal configuration depends on users' desired latency and cost targets, 
(c) input video contents may exercise different paths in the DAG and produce a variable amount intermediate results.
Existing video analytics and processing systems leave it to the users to manually configure operations and select hardware resources.

We present \SystemName: a heterogeneous and serverless framework for auto-tuning video pipelines.
Given an end-to-end latency target, \SystemName optimizes for cost efficiency by (a) calculating a latency target for each operation invocation, and (b) dynamically running a cost-based optimizer to assign configurations across heterogeneous hardware that best meet the calculated per-invocation latency target.
This makes the problem of auto-tuning large video pipelines tractable and allows us to handle input-dependent behavior, conditional branches in the DAG, and execution variability.
We describe the algorithms in \SystemName and evaluate it on a cloud platform using serverless CPU and GPU resources.
We show that compared to state-of-the-art cluster and serverless video analytics and processing systems, \SystemName achieves 7.8$\times$ lower latency and 16$\times$ cost reduction on average. 
\end{abstract}

\maketitle
\pagestyle{plain} %

\section{Introduction}
\label{sec:introduction}
Video traffic is exploding in scale, predicted to account for over 82\% of all internet traffic by 2022~\cite{cisco-internet-whitepaper}.
A myriad of domains use \textit{video pipelines}, with tens of video analytics and processing operations, to extract meaningful information from raw videos.
For example, an AMBER Alert application can leverage traffic cameras across a city to pinpoint specific individuals and cars~\cite{amber-alerts}.
To do so, the application uses a pipeline to first detect frames with people and/or cars, and then match them to specific individuals' faces and car descriptions, respectively.
As video analytics research continues to flourish, we expect a perpetual proliferation of emerging domains that depend on video pipelines, such as smart cities~\cite{nsdi17-videostorm, sigcomm18-chameleon, osdi18-focus, computer17-video-analytics}, surveillance analytics~\cite{surveillance}, healthcare~\cite{healthcare}, and retail~\cite{amazongo}.

The pervasive use of video analytics applications has led to significant challenges.
Video pipelines must meet a wide range of latency, throughput, and cost targets to be practical across applications.
For example, a pipeline to detect cars and people in a traffic feed should be tuned to be more cost efficient for city traffic planners with relaxed latency targets, while the same pipeline must be tuned to meet strict latency targets for AMBER Alert responders~\cite{amber-alerts}.
Video analytics and processing frameworks must tune pipeline operation knobs (e.g., sampling rate, batch size, hardware target, and resource allocation) to meet the unique latency or cost requirements of diverse applications.
However, automatically tuning these knobs is difficult for the following reasons.

\myparagraph{Operations exhibit performance variation across heterogeneous hardware.}
Hardware accelerators (e.g., GPUs~\cite{hotchips-gpu}, FPGAs~\cite{hotchips-fpgaintel,hotchips-fpgaxilinx}, TPUs~\cite{isca17-tpu}, and vision chips~\cite{ambarella}) provide significant performance benefits for many pipeline operations.
Tuning knobs across these heterogeneous accelerators can have a huge impact in the performance and efficiency of video pipelines.
We observed a 3.7$\times$ latency variation by tuning CPU cores, GPU memory, and batch size for operations in a representative AMBER Alert pipeline processed using Scanner \cite{siggraph18-scanner}.
While recent research has proposed mechanisms to tune operation knobs based on resource usage~\cite{sigcomm18-chameleon, osdi18-focus, atc18-videochef}, they are limited to simple pipelines and homogeneous hardware platforms.
Furthermore, they rely on hours to days of profiling for \emph{each} new pipeline, video, and latency target~\cite{nsdi17-videostorm,nsdi17-cherrypick}.

\myparagraph{Pipelines can have input-dependent execution flow. }
An input video's contents influence the execution flow of a pipeline in two ways.
First, the number of intermediate outputs for an operation may depend on the frame being processed.
For the AMBER Alert pipeline, the object detector operation will output a variable number of cropped people and/or car images to be processed by subsequent operations.
Second, downstream operations may be conditionally executed based on the intermediate output.
For example, if there are only people in a frame, no car classification is needed.
Consequently, tuning configuration knobs and resource allocations dynamically based on video content is critical for performance and efficiency.
We found the static configurations made by gg~\cite{atc19-gg}, a general purpose serverless framework, degraded performance by as much as 57\% for the AMBER Alert pipeline. 
Some systems, such as VideoStorm~\cite{nsdi17-videostorm} and GrandSLAm~\cite{eurosys19-grandslam}, only support simple sequential pipelines with deterministic flow. 

Systems that use serverless platforms as their backend (e.g., ExCamera~\cite{nsdi17-excamera}, gg~\cite{atc19-gg}, PyWren~\cite{socc17-pywren}, and Sprocket~\cite{socc18-sprocket}) execute applications by using thousands of short-lived functions~\cite{awslambda, gcf, azurefunctions}.
The function-level resource allocation offered by serverless platforms makes them an attractive option for processing video pipelines, as they enable dynamic tuning for each operation invoked.
However, existing serverless offerings lack support for heterogeneous hardware accelerators and application constraints such as latency targets.
Users must still manually, and perhaps exhaustively, explore operation knobs and resource allocation options.

We present \SystemName, a video analytics and processing framework that supports heterogenous hardware and automatically tunes each operation invocation to meet diverse latency targets for the overall pipeline.
\SystemName is a full-fledged serverless framework that provides a serverless experience to its users, who do not need to express the resources or operation knob configurations needed to meet their latency targets. %

\SystemName is divided into two parts: an offline specification phase, and an online optimizaiton phase.
The offline specification phase allows users to specify their pipeline, and performs a one-time, per-operation profiling that allows \SystemName to automatically tune operation invocations as the pipeline runs.
Unlike existing systems~\cite{socc20-inferline}, this profiling does not need to be repeated as the pipeline or input video changes.

During the online phase, \SystemName leverages two key ideas to meet diverse latency targets.
First, \SystemName \emph{dynamically} computes how much time can be spent on each invocation to meet the pipeline latency target (i.e., per-invocation latency targets).
By computing a per-invocation latency target, \SystemName can dynamically explore the configuration space for each invocation and adapt to performance volatility and input-dependent execution flows.
Second, \SystemName dynamically runs a cost-based optimizer that determines the most efficient operation configuration that meets the per-invocation target.
To do so, \SystemName
(a) uses \emph{early speculation and late commit}: a technique for choosing an initial operation knob configuration during pipeline processing, and revisiting the configuration right before execution,
(b) leverages \emph{priority-based commit} to prioritize operations based on hardware affinity and DAG dependencies, and 
(c) employs \emph{safe delayed batching} to batch operations for efficiency as long as doing so does not violate per-invocation targets.

We deploy \SystemName on Google Cloud Platform with serverless CPU and GPU backends and evaluate its efficiency and ability to meet latency targets for five video analytics and processing pipelines. 
By dynamically configuring operations for both CPU and GPU based on pipeline latency targets, \SystemName achieves on average 7.8$\times$ latency improvement and 16$\times$ cost reduction compared to four state-of-the-art cluster and serverless video analytics and processing systems: Nexus~\cite{sosp19-nexus}, Scanner~\cite{siggraph18-scanner}, gg~\cite{atc19-gg}, and GrandSLAm~\cite{eurosys19-grandslam}.

\section{Background and Motivation}\label{sec:motivation}

\begin{figure}[t]
\centering
  \begin{minipage}[t]{0.46\linewidth}
    \centering
    \includegraphics[width=1.0\linewidth]{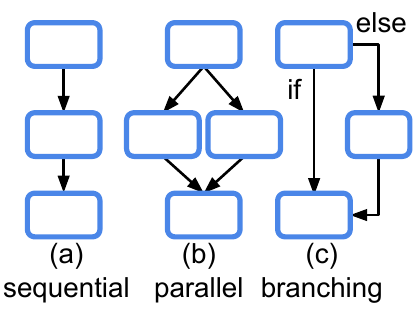}
    \caption{\small Simple DAGs that can be used to compose complex video pipelines.}
    \label{fig:build-blocks}
  \end{minipage}
  \hfill
  \begin{minipage}[t]{0.46\linewidth}
    \centering
    \includegraphics[width=1.0\linewidth]{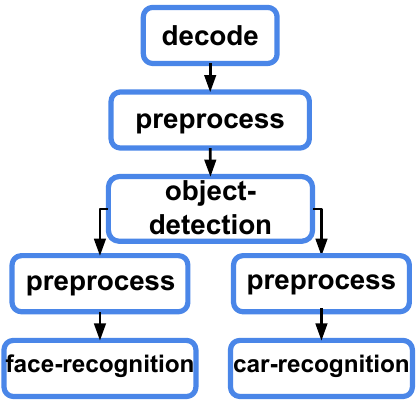}
    \caption{\small An AMBER Alert pipeline that finds faces and cars in a video.}
    \label{fig:example-pipeline}
  \end{minipage}
\end{figure}

\begin{table*}[t]
  \centering
  \footnotesize
  \resizebox{\linewidth}{!}{%
  \begin{tabular}{l|l|lllllllll}
            &         & \multicolumn{1}{c}{\begin{tabular}[c]{@{}c@{}}InferLine\\ \cite{socc20-inferline}\end{tabular}} & \multicolumn{1}{c}{\begin{tabular}[c]{@{}c@{}}GrandSLAm\\ \cite{eurosys19-grandslam}\end{tabular}} & \multicolumn{1}{c}{\begin{tabular}[c]{@{}c@{}}VideoStorm\\ \cite{nsdi17-videostorm}\end{tabular}} & \multicolumn{1}{c}{\begin{tabular}[c]{@{}c@{}}Focus\\ \cite{osdi18-focus}\end{tabular}} & \multicolumn{1}{c}{\begin{tabular}[c]{@{}c@{}}Nexus\\ \cite{sosp19-nexus}\end{tabular}} & \multicolumn{1}{c}{\begin{tabular}[c]{@{}c@{}}Scanner\\ \cite{siggraph18-scanner}\end{tabular}} & \multicolumn{1}{c}{\begin{tabular}[c]{@{}c@{}}gg\\ \cite{atc19-gg}\end{tabular}} & \multicolumn{1}{c}{\begin{tabular}[c]{@{}c@{}}Sprocket\\ \cite{socc18-sprocket}\end{tabular}} & \textbf{\SystemName}    \\ \hline

& Performance targets                & {\color[HTML]{32CB00} Yes} & {\color[HTML]{32CB00} Yes}     & {\color[HTML]{32CB00} Yes}     & {\color[HTML]{32CB00} Yes}      & {\color[HTML]{32CB00} Yes}     & {\color[HTML]{FE0000} No}  & {\color[HTML]{FE0000} No}      & {\color[HTML]{FE0000} No}  & {\color[HTML]{32CB00} Yes} \\
\multirow{-2}{*}{\textbf{Features}}   & General operations                 & {\color[HTML]{FE0000} No} & {\color[HTML]{FE0000} No}      & {\color[HTML]{32CB00} Yes}     & {\color[HTML]{FE0000} No}      & {\color[HTML]{FE0000} No}      & {\color[HTML]{32CB00} Yes} & {\color[HTML]{32CB00} Yes}     & {\color[HTML]{FE0000} No}  & {\color[HTML]{32CB00} Yes} \\ \hline
& Traverse large configuration space & {\color[HTML]{CD9934} Limited\textsuperscript{$\mathparagraph$}} & {\color[HTML]{CD9934} Limited\textsuperscript{$\mathparagraph$}} & {\color[HTML]{CD9934} Limited\textsuperscript{$\dagger$}} & {\color[HTML]{CD9934} Limited\textsuperscript{$\mathparagraph$}} & {\color[HTML]{CD9934} Limited\textsuperscript{$\mathparagraph$}} & {\color[HTML]{FE0000} No}  & {\color[HTML]{FE0000} No}      & {\color[HTML]{FE0000} No}  & {\color[HTML]{32CB00} Yes} \\
& Handle input-dependent exec. flow & {\color[HTML]{FE0000} No} & {\color[HTML]{FE0000} No}      & {\color[HTML]{FE0000} No}      & {\color[HTML]{FE0000} No}      & {\color[HTML]{32CB00} Yes}     & {\color[HTML]{FE0000} No}  & {\color[HTML]{CD9934} Limited\textsuperscript{$\ddagger$}} & {\color[HTML]{FE0000} No}  & {\color[HTML]{32CB00} Yes} \\
\multirow{-3}{*}{\textbf{Challenges}} & Dynamically adjust resource alloc. & {\color[HTML]{32CB00} Yes} & {\color[HTML]{FE0000} No}      & {\color[HTML]{CD9934} Limited\textsuperscript{$\mathsection$}}      & {\color[HTML]{FE0000} No}      & {\color[HTML]{CD9934} Limited\textsuperscript{$\mathsection$}} & {\color[HTML]{FE0000} No}  & {\color[HTML]{32CB00} Yes}     & {\color[HTML]{32CB00} Yes} & {\color[HTML]{32CB00} Yes}
  \end{tabular}
  } %
  \caption{\small
    Comparison of existing video processing systems with \SystemName based on whether they (a) support performance targets and general operations, and (b) address the challenges of meeting performance targets for general video pipelines.
    $\mathparagraph$Limited to domain-specific knobs. $\dagger$Large profiling overhead. $\ddagger$Cannot handle branching. $\mathsection$Limited to single hardware platform.
  }
  \label{tab:compare-features}
  \vspace{-6mm}
\end{table*}

Applications define \emph{video pipelines} as directed acyclic graphs (DAGs), where vertices represent video analytics and processing operations, while edges represent dataflow. 

As described in literature~\cite{aws-step-functions, eurosys19-grandslam, sec18-costless}, video pipelines can be composed from three basic DAG patterns shown in Figure~\ref{fig:build-blocks}: (a) sequential, where each vertex has at most one input and one output, (b) parallel, where multiple vertices execute in parallel, and (c) branching, where the output of a vertex, called branching vertex, conditionally determines the next vertex to execute.
For example, the AMBER Alert pipeline~\cite{nsdi17-videostorm, sosp19-nexus} for face and car recognition in Figure~\ref{fig:example-pipeline} begins with a sequential path of decoding and preprocessing operations, followed by a branching object detection operation.
Depending on the output, people or cars are sent to parallel face and car recognition operations, respectively.

Table ~\ref{tab:compare-features} categorizes video analytics and processing systems based on two key features: 
(a) Their ability to specify and meet \textit{performance targets.} User-facing systems typically require that the video pipeline meet a latency target, ideally while minimizing resource usage (cost).
For example, the AMBER Alert pipeline needs to meet a strict latency target so that responders can take timely action.
(b) Support for \textit{general video operations.} To compose video pipelines, a user combines video operations (e.g., inference models, video encoders, and image filters) with analytics operations that process extracted metadata.
For example, the AMBER Alert pipeline will contain video decoding, object detection, face recognition, and car model recognition.
Some systems, such as Scanner~\cite{siggraph18-scanner}, VideoStorm~\cite{nsdi17-videostorm}, and gg~\cite{atc19-gg}, support general video operations.
Others, such as Focus~\cite{osdi18-focus}, Nexus~\cite{sosp19-nexus}, GrandSLAm~\cite{eurosys19-grandslam}, and InferLine~\cite{socc20-inferline} focus on one facet of video pipelines (e.g., deep learning inference) and rely on external services for other operations.

\subsection{Challenges} \label{sec:challenges}
\myparagraph{Large configuration space. }
Pipeline operations offer a variety of \textit{knobs} that can be tuned to improve latency and resource use. 
For example, many operations have knobs such as batch size, sampling rate, and resolution.
Other knobs select the hardware platform (e.g., CPU, GPU, TPU, etc.) and set the resource allocation (e.g., CPU cores or GPU memory).
Determining configurations is challenging due to the exponential growth in the configuration space with the number of operations, knobs, and hardware platforms available.

As shown in Table~\ref{tab:compare-features}, Scanner, gg, and Sprocket do not auto-tune configurations knobs, putting the burden on the user to statically specify good operation configurations.
Focus, Nexus, GrandSLAm, and InferLine are domain-specific to deep learning inference and are limited to configuring the inference models used and the batch size.
VideoStorm supports general knob configurations; however, it takes tens of CPU hours to profile pipelines and requires re-profiling when the pipeline, input video, or latency targets change~\cite{nsdi17-videostorm}.

\myparagraph{Input-dependent execution flow. }
\begin{figure}[t]
\centering
  \begin{subfigure}[t]{0.488\linewidth}
      \centering
      \includegraphics[width=1.0\linewidth]{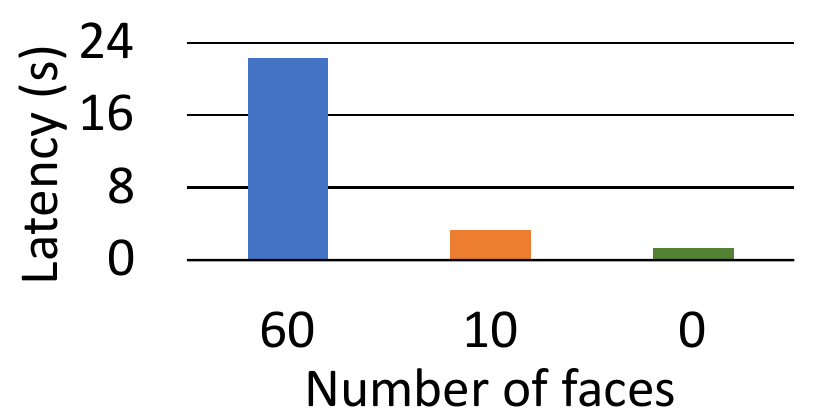}
  \end{subfigure}
  \hfill
  \begin{subfigure}[t]{0.49\linewidth}
      \centering
      \includegraphics[width=1.0\linewidth]{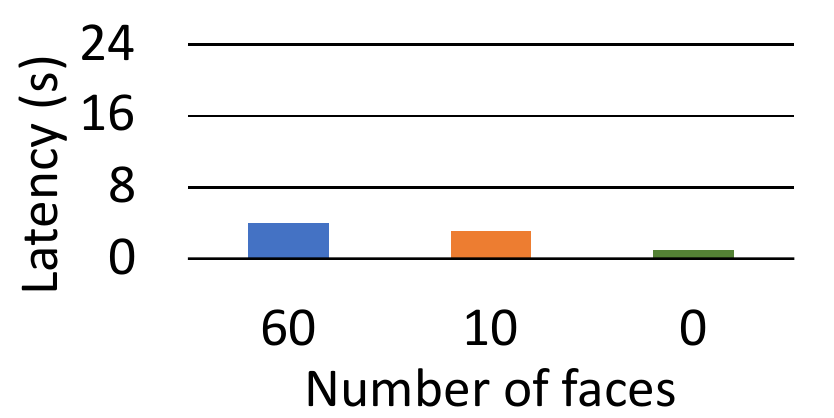}
  \end{subfigure}
  \vspace{-2mm}
  \caption{\small Execution latency on CPU (left) and GPU (right) for a face detection pipeline that identifies unique faces in a frame~\protect\cite{dlib09}. Latency varies up to 17.2$\times$ and 4$\times$ on CPU and GPU, respectively.} 
  \vspace{-3mm}
  \label{fig:facerecog}
\end{figure} 

Input-dependent execution flow occurs in two cases: 
First, inputs determine the conditional path in branching pipelines.
In the AMBER Alert pipeline of Figure~\ref{fig:example-pipeline}, a frame will only take the face recognition path if \texttt{object\-/detection} finds a person in it.
Since a conditional path is not resolved until the branching operation finishes, provisioning resources and selecting configurations to meet a pipeline's latency target is challenging.
Existing systems either treat branching pipelines as parallel ones (i.e., by executing all conditional branches)~\cite{siggraph18-scanner,eurosys12-jockey,socc20-inferline} or do not support non-sequential pipelines~\cite{eurosys19-grandslam,nsdi17-videostorm,socc18-sprocket}.

Second, operations can produce a variable number of outputs, and thus a variable load for downstream operations. 
If the number of intermediate outputs is unknown until the operation is executed, determining the parallelism or resources needed downstream to meet latency targets is difficult, especially if these operations are computationally expensive.  
Figure~\ref{fig:facerecog} shows the latency for a pipeline that identifies the unique faces in a frame depends on the number of unique faces: 17.2$\times$ and 4$\times$ difference between 60 faces versus no faces on a CPU and a GPU, respectively.
Thus, the nondeterminism introduced by input-dependent behavior requires systems to dynamically adapt to meet a pipeline's latency target~\cite{eurosys12-jockey}.
Most existing video analytics and processing systems do not account for input-dependent execution flow.

\myparagraph{Dynamically adjusting resource allocation of operation invocations. }
As a pipeline executes, the degree of available parallelism depends on the various knob settings (e.g., batching) and the number of intermediate outputs.
Many existing systems require users to statically provision a cluster, which limits the resources available to exploit parallelism~\cite{eurosys19-grandslam,eurosys12-jockey} or leads to over-allocation and higher costs when parallelism is low.
Some systems periodically adjust resources and bin pack requests as the load changes, but are limited by how quickly hardware (e.g., GPUs) and VMs can be loaded/unloaded~\cite{sosp19-nexus}.
Systems like gg~\cite{atc19-gg} and Sprocket~\cite{socc18-sprocket} leverage serverless platforms~\cite{awslambda,azurefunctions,gcf} to dynamically allocate resources for each operation invocation.
However, serverless platforms still require users to manually select hardware types and configure knobs to meet latency targets.

\section{\SystemName Design} \label{sec:design}
\SystemName is a heterogeneous and serverless framework for auto-tuning video analytics and processing pipelines.
\SystemName's objective is to meet the overall pipeline latency target, while minimizing cost (resource usage). 
As noted in Section~\ref{sec:motivation}, input-dependent execution flow and resource volatility preclude the use of static tuning approaches~\cite{eurosys12-jockey}.
They also preclude designing and calculating a globally-optimal solution a-priori or dynamically.
Instead, \SystemName optimizes the overall pipeline execution by iteratively and dynamically optimizing each operation invocation using the most up-to-date information about the state of execution flow and resource availability.
Specifically, \SystemName (a) dynamically reduces the pipeline target latency to per-operation invocation latency targets, values that we call \emph{slack}, and (b) continuously configures each operation invocation to meet the slack at minimal cost.
Dynamically allotting slack ensures the pipeline latency target is met without having to statically account for all possible conditional paths or sources of resource volatility in serverless environments.
It also allows \SystemName to revisit configuration decisions as the resource environment evolves or as input-dependent operations are run.
\SystemName finds the set of cost-efficient configurations for the entire pipeline because it minimizes cost at each configuration assignment subject to the overall latency target.

We address the challenges outlined in Section~\ref{sec:motivation} as follows:

\myparagraph{Traversing the large configuration space. }
\SystemName profiles and makes configuration decisions on a \textit{per-operation}, not per-pipeline basis.
New operations undergo a short (seconds to minutes), one-time profiling step independent of the pipelines that include the operation.
Operations are not re-profiled as the pipeline composition, video, or latency targets change.
As the pipeline executes, \SystemName makes configuration decisions for one operation invocation at a time, reducing the exponential configuration space of an entire pipeline to that of an individual operation. 

\myparagraph{Handling input-dependent execution flow. }
\SystemName uses three techniques to meet latency targets despite the nondeterminism that stems from input-dependent behavior and resource volatility (e.g., resource contention):
(a) \emph{early speculation and late commit} selects an initial configuration decision as soon as an invocation is available, then revisits the configuration right before execution,
(b) \emph{priority-based commit} prioritizes operations based on their affinity to hardware and their depth in the pipeline, and 
(c) \emph{safe delayed batching} waits for additional inputs for batching, as long as doing so does not violate the invocation's allotted slack.

\myparagraph{Dynamically adjusting resource allocations. }
Making per-invocation configuration decisions also allows \SystemName to dynamically right-size resource allocations across heterogeneous serverless backends.
\SystemName decides the hardware type and resource sizing (e.g., GPU with 2GB of memory) during dynamic configuration based on what is necessary to meet the slack.
Early speculation and late commit, as well as priority-based commit, also allow \SystemName to balance resources between operations. 

\subsection{Architecture}

\begin{figure}[t]
 \centering
 \includegraphics[width=.95\columnwidth]{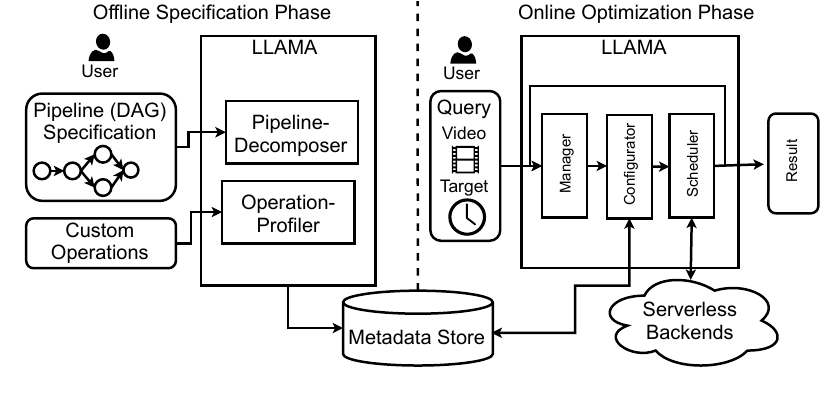} 
 \vspace{-4mm}
 \caption{\small \SystemName's architecture diagram.}
 \label{fig:block-diagram}
 \vspace{-3mm} 
\end{figure}

\SystemName uses an offline specification phase and an online optimization phase (Figure~\ref{fig:block-diagram}).
The specification phase has two purposes.
First, it allows the user to specify a pipeline with multiple, general operations using a SDK.
Second, it extracts the following information: a set of all possible sequential paths through the pipeline, and the latency and resource footprint of each unique operation across possible knob configurations.  The pipeline specification and the extracted metadata are stored for use during the online phase.

The online phase is triggered when users submit an input video and a latency target to \SystemName.
\SystemName executes the pipeline by continuously generating and executing a set of invocations for each operation as their input dependencies are resolved.
For example, if \texttt{object-detection} outputs a frame tagged with a person, a new invocation is generated for the \texttt{preprocess} operation in the AMBER Alert pipeline (Figure~\ref{fig:example-pipeline}).
The online phase configures each invocation by first estimating its slack.
It then uses the respective operation's profiling data to determine the most efficient configuration for completing the invocation within the allotted slack.
The process repeats until all pipeline invocations are executed.

\subsection{Specification phase} \label{sec:offline}
\myparagraph{Application Interface. }
Users specify pipeline operations, dependencies between operations, and conditional flow using the \SystemName SDK.
\SystemName provides a library of operations (e.g., decode and face recognition).
Each operation consists of a binary executable, indexed by its SHA256 hash, and a \textit{configuration specification} file that contains configuration options and performance statistics for the operation.
Users can optionally bring their own operations by providing an executable and a configuration template that specifies tunable knobs (e.g., hardware type, batch size, or number of filters), the ranges for each knob, and the granularity of this range (e.g., batch size increasing by powers of 2).
The Operation-Profiler uses these inputs in a one-time profiling step to generate a configuration specification. 
The operation and configuration specification are then added into the Metadata Store and re-used across pipelines without further profiling. 

\myparagraph{Operation-Profiler. }
The Operation-Profiler collects performance and resource statistics for each operation.
Using the operation executable and configuration template as inputs, it first enumerates all possible configurations specified by the template, then executes a short profiling step using one or more sample frames for each configuration (depending on the batch size).
Statistics such as latency and resource footprint (e.g., peak memory utilization) are collected and stored as configuration specification file entries.
The frame content does not affect these statistics (recall that input-dependent execution flow manifests between operations). 
Since slack calculation (Section \ref{sec:slack}) is only dependent on the \emph{relative} performance of operation invocations across the pipeline, the Operation-Profiler designates a \emph{reference configuration} for each operation to provide a measure of relative performance.
We chose the smallest CPU configuration (1-core, batch-1) for each operation's reference configuration.
During runtime, operation invocation performance that differs from its profiled value, due to resource contention or profiling inaccuracy, is managed by leveraging feedback (Section~\ref{sec:onlinepreview}).

The configuration specification is structured so arbitrary operation- and hardware-specific configuration knobs can be described by users, and dynamically configured during runtime.
This enables \SystemName to support general operations and arbitrary video pipelines for a myriad of applications.

\myparagraph{Pipeline-Decomposer. }
To enable the online phase to dynamically compute slack, the Pipeline\-/Decomposer performs a one-time \emph{decomposition} of the pipeline into all possible sequential paths in the pipeline.
To do so, it performs a modified depth-first search on the pipeline DAG to enumerate all paths from the input operation (i.e., operation with no upstream dependencies) to an output operation (i.e., an operation with no downstream dependencies).
It then emits an intermediate representation of the decomposed paths into the Metadata Store.
For example, the AMBER Alert pipeline in Figure~\ref{fig:example-pipeline} is decomposed into the two sequential paths ending in \texttt{face\-/recognition} and \texttt{car\-/recognition}, respectively.

\subsection{Online phase}\label{sec:onlinepreview}

\myparagraph{Manager. }
\SystemName's Manager takes video inputs and latency targets and orchestrates the entire pipeline execution, maintaining execution state and generating new invocations.
Whenever an invocation completes, the Manager records the invocation's runtime statistics (i.e., latency, cost, and configuration) and the location of intermediate outputs.
The runtime statistics are used to update the configuration profiles obtained from the Operation-Profiler via a feedback loop.
We use an exponential smoothing algorithm to update the profiling; other algorithms can be incorporated as well.
The intermediate outputs are then used to resolve any conditional branches. The Manager then spawns invocations for downstream operations once all dependencies have been resolved. 
These invocations are then sent to the Configurator. 

\begin{figure}[t]
 \centering
 \includegraphics[width=.95\columnwidth]{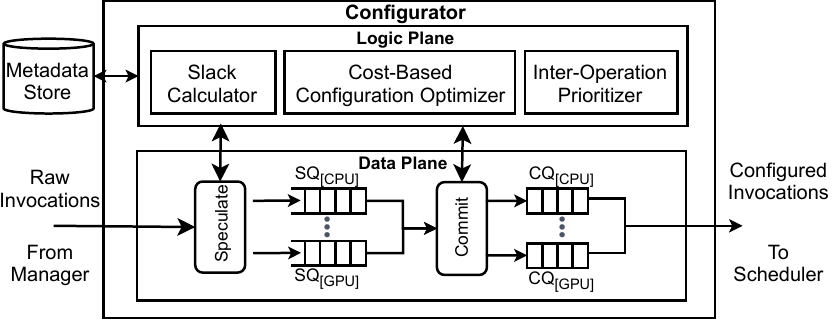} 
 \caption{\small Configurator diagram.}
 \label{fig:configurator-diagram}
\end{figure}

\myparagraph{Configurator. }
To meet the overall pipeline latency target, the Configurator (Figure~\ref{fig:configurator-diagram}) decides (a) how much slack to allot to an operation invocation, and (b) what the most efficient configuration is to meet the slack.
The Configurator works with the Scheduler to keep track of available resources at a serverless backend as it makes configuration decisions.

\myparagraph{Scheduler. }
After the Configurator has configured an invocation's knobs, the invocation is sent to the Scheduler for execution.
The Scheduler executes the configured invocations on the hardware platform specified by the Configurator.
This includes creating and managing the necessary backend connections, mitigating stragglers, and handling invocation failures (Section~\ref{sec:fault-tol}).
When an invocation successfully returns, the Scheduler provides the Manager with the invocation metadata and output results.

\section{Target Latency-Aware Configuration}\label{sec:online}
Input-dependent execution flow and backend resource volatility require the Configurator to dynamically determine each operation invocation's most efficient knob configurations.
The Configurator is divided into two parts.
The Logic Plane (a) determines how much slack can be spent on its invocation, and (b) uses a cost-based optimizer to select a configuration to meet that slack.
The Data Plane manages configured operations in queues prior to their execution.

\begin{algorithm}[t]
  \caption{Operation invocation slack allotment}\label{alg:slack}
  \begin{algorithmic}[1]
\State $paths \gets$ A set of all sequential paths in the pipeline
\State $t \gets$ elapsed time
\State $target \gets$ pipeline latency target %
\Procedure{ComputeSlack}{$op, \lambda$}
  \State $slacks = \{\}$
  \ForAll{\{$path \in paths \mid op \in path$\}}
    \State $pLat = \textsc{RemainingPathLatency}(op, path)$
    \State $remainingTime = (target - t - queueingTime(\lambda))$
    \State $pSlack = (op.ReferenceLat()/pLat) \times remainingTime$
    \State $slacks.append(pSlack)$
  \EndFor
  \State \textbf{return} $min(slack \in slacks)$
\EndProcedure
\end{algorithmic}
\end{algorithm}
 
\subsection{Determining an invocation's slack}\label{sec:slack}  
Given a user-specified pipeline latency target, the Configurator first needs to compute a slack for each operation invocation.
Existing systems (e.g., GrandSLAm~\cite{eurosys12-jockey} and Fifer~\cite{middleware20-fifer}) statically determine each operation invocation's slack by assuming a linear pipeline with predictable invocations and latencies (i.e., no nondeterminism).
Our insight is to instead \emph{dynamically} calculate each operation invocation's slack, which we subsequently use to select the best invocation configuration (Section~\ref{sec:objective}).
Doing so across invocations efficiently meets the pipeline latency target.

\SystemName calculates an operation invocation's slack using Algorithm~\ref{alg:slack}.
Given an invocation of operation $op$ and a configuration's backend $\lambda$, \textsc{ComputeSlack} begins by finding every sequential path through the DAG containing $op$ (Line 6).
It then estimates the latency to complete the path, starting at $op$, using the reference configuration for each operation (Section~\ref{sec:offline}).
By using the reference configuration's latency, \SystemName avoids a causal dilemma of needing a configuration to compute slack, and needing a slack to select a configuration.
The operation invocation's slack for that path is then determined based on the remaining time (Line 8), factoring in estimated queueing time at $\lambda$, weighted by the relative latency of $op$ to the remaining path (Line 9).
The final slack for an invocation of $op$ on $\lambda$ is then the minimum slack value over all possible execution paths of $op$, which accounts for all input-dependent branch resolutions (Line 11).
We discuss how \SystemName reclaims overly-conservative slack next.

\subsection{Navigating the configuration space}\label{sec:objective} 
Since slack is calculated for each operation invocation, \SystemName can quickly evaluate configurations in a smaller per-operation, not per-pipeline, configuration space.
After calculating the invocation's slack for each available serverless backend $\lambda$ (Algorithm~\ref{alg:slack}), 
\SystemName applies the objective function shown in Equation~\ref{eq:objective} for all possible configurations $x$ of $op$ using the invocation slack corresponding to the serverless hardware backend $\lambda(x)$ targeted by $x$.
$C(x)$ and $L(x)$ are the estimated cost and latency respectively to run configuration $x$.
$R(x)$ is the resources requested by $x$ (e.g., amount of memory), and $R_{total}(\lambda(x))$ is the resource limit of $\lambda(x)$.
$B(x)$ is the batch size of configuration $x$, and $\alpha$ is a tunable weight.

\begin{equation} \label{eq:objective}
  \small
  obj(x, slack) = \begin{cases}
    C(x) / B(x) & L(x) < slack \\
    \frac{C(x)}{B(x)} + \alpha\frac{(L(x) * R(x))}{(B(x) * R_{total}(\lambda(x))} & otherwise
  \end{cases}
\end{equation}

Intuitively, this objective function evaluates the monetary cost to run $x$ when there is a feasible slack.
If slack cannot be met (e.g., if the user submits an unachievably low target), the cost function weighs in favor of potentially more expensive configurations that achieve a higher throughput.
$\alpha$ sets the balance between cost and throughput, with high values of $\alpha$ set to meet the slack at all costs, while lower values of $\alpha$ may leverage more cost-efficient configurations potentially at the expense of exceeding slack.
The configuration objective function is independent of the input video or overall pipeline.

Users who wish to optimize for a different metric (e.g., minimal latency subject to a cost budget) can add their own objective function to \SystemName.
Furthermore, since $R$ is specific to each backend (e.g., concurrent invocation limits, GPU memory, or CPU cores), \SystemName can support other heterogeneous backends (e.g., serverless GPUs or on-premise clusters).

Since conditional flow will not always resolve to the worst-case path, the allotted slack may result in a configuration with a lower-than-necessary latency.
However, since each invocation is configured separately and dynamically, future invocations will recover efficiency from earlier mis-predictions.

\subsection{Revisiting configuration decisions} \label{sec:queueing} 
To manage invocations that cannot be run concurrently due to limited backend parallelism, \SystemName locally queues invocations and accounts for the queueing time when allotting slack (Line 9 in Algorithm~\ref{alg:slack}).
The queueing time depends on $x_i$: the selected configuration of each queued invocation $i$ (i.e., it is not sufficient to use the number of queued operation invocations as a measure of wait time~\cite{eurosys12-jockey,eurosys16-tetrisched}).
Thus, invocations need to be assigned a configuration before they are queued.
However, the initial configuration $x_i$ is often made many seconds before it is actually invoked, leading to sub-optimal configurations for three reasons.
(a) Invocations queued \textit{in front of} $i$ may experience execution times that vary from the profiled values.
This can occur due to resource contention or input-dependent execution flow.
(b) The estimated latency for $x_i$ may be updated via feedback while it is queued.
(c) The number of invocations queued \textit{behind} $i$ may quickly grow (e.g., many completed \texttt{object\-/detection} invocations may output a large number of \texttt{car\-/recognition} and \texttt{face\-/recognition} invocations); thus, $x_i$ should be chosen to ensure upstream invocations can meet their slack.
Hence, by the time a queued invocation is ready to run, its selected configuration needs to be revisited to determine if it is still the right configuration.

To solve this, \SystemName leverages a novel technique inspired by late binding \cite{nsdi20-sol,sosp13-sparrow,eurosys18-bindschaedler, osdi18-qoop, osdi14-venkataraman} that we call \textbf{early speculation and late commit}.
With early speculation and late commit, \SystemName maintains two queues per serverless backend $\lambda$: an unbounded speculative queue ($SQ[\lambda]$) and a small, bounded commit queue ($CQ[\lambda]$) set to hold enough invocations to saturate $\lambda$.
Once an invocation $i$ is ready to execute, the Configurator uses Algorithm~\ref{alg:slack} to assign it a slack, and uses Equation~\ref{eq:objective} to select a \emph{speculative configuration}.
The configured invocation is then put into the appropriate speculative queue, thus enabling \SystemName to estimate the queueing time at each backend.
Once $i$ reaches the head of the speculative queue, as prior invocations are executed, \SystemName revisits the configuration of $i$ by using Algorithm~\ref{alg:slack} and Equation~\ref{eq:objective} again.
It then \emph{commits} the configuration into the appropriate commit queue.
Doing so mitigates the queueing challenges we noted above by delaying binding an invocation to a final configuration for as long as possible.
This provides \SystemName with maximum flexibility and the most up-to-date state about pipeline dataflow and performance at each backend.

With early speculation and late commit, \SystemName can estimate the queueing time using Equation~\ref{eq:queueing} for each serverless backend $\lambda$ based on each configured invocation $i$ in its queues.
$L(x_i)$ and $R(x_i)$ are the estimated latency of, and resources requested by $x_i$, respectively.
$R_{total}(\lambda(x_i))$ is the total amount of resources or concurrency limit at the serverless backend specified by the configuration $x_i$.

\begin{equation}\label{eq:queueing}
  \small
  Q_{SQ[\lambda],CQ[\lambda]} = \sum_{i \in \{SQ[\lambda], CQ[\lambda]\}} L(x_i) \frac{R(x_i)}{R_{total}(\lambda(x_i))}
\end{equation}

Intuitively, the queueing time is the sum of each $x_i$'s profiled configuration latency, weighted by $x_i$'s requested resources (to account for parallel execution).
The cumulative queueing time over $SQ[\lambda]$ and $CQ[\lambda]$ is then used in \textsc{ComputeSlack}.
$SQ[\lambda]$ is included when committing configured invocations to account for invocations queued behind $i$.
We do not incorporate future operations' queueing time, since dynamicity and the need to assign a configuration to each downstream operation can result in inaccurate estimates.

\subsection{Inter-operation prioritization}\label{sec:inter-operation}
\subsubsection{Challenges}~\\
The Configurator's decisions described in Section~\ref{sec:objective} assume per-operation invocation decisions can be made independently of each other.
However, \SystemName also needs to reason about the relationship between operations and their invocations for the following reasons:

\myparagraph{When to batch invocations. }
As pipeline dataflow progresses, there can be moments when an operation may have fewer invocations available than the most efficient configuration's batch size.
For example, if a pipeline contains a slower face detection operation followed by a faster blurring operation, the blur operation's invocations will likely drain the speculative queue faster than it can build up.
In such cases, executing upstream operations first yields a larger batch size, amortizing RPC and I/O overheads.
However, waiting for upstream operation invocations to complete their execution may result in a slack violation.

\myparagraph{Under-allotting slack due to incorrect profiling. }
As described in Section~\ref{sec:slack}, slack allotted to an invocation is a function of the reference configuration's profiled latency for downstream operations.
Furthermore, a configuration's latency is updated using a feedback loop after execution (Section~\ref{sec:onlinepreview}).
However, slack can be under-allotted to operation invocations if the reference configuration latency is significantly shorter than its actual latency, and the feedback loop is not closed early on during pipeline execution.
This is especially problematic for longer pipelines, and for pipelines with the last operation's invocations needing a longer slack than the rest.
Thus, it is beneficial to prioritize invocations by pipeline depth early in the pipeline's execution so that feedback can update all reference configurations.

\myparagraph{Affinity of operations to heterogeneous hardware. }
While prioritizing invocations by pipeline depth can help prevent under-allotting slack, the issue of prioritizing operation invocations on particular hardware platforms still remains. 
For example, consider the case in which both an \texttt{object\-/detection} and \texttt{face\-/recognition} invocation must be configured.
Assume that while both operations run faster on a GPU, \texttt{face\-/recognition} benefits more from acceleration and observes a larger latency reduction.
Resource limits force the two invocations to split their decision between $\lambda_{CPU}$ and $\lambda_{GPU}$.
Committing \texttt{object\-/detection}'s invocation first forces \texttt{face\-/recognition}'s invocation to choose $\lambda_{CPU}$.
However, the better decision is to assign the CPU to \texttt{object\-/detection} and the GPU to \texttt{face\-/recognition}.
The relative benefit of running operation invocations on a particular hardware platform (i.e., its hardware affinity) must be incorporated into configuration decisions.

\subsubsection{Our solution}~\\
\SystemName addresses these challenges using \textbf{safe delayed batching} and \textbf{priority-based commit}, implemented in conjunction with early speculation and late commit.

\myparagraph{Safe delayed batching. }
Safe delayed batching addresses the challenge of waiting for additional invocations to batch.
During both early speculation and late commit, if \SystemName determines the most cost-efficient configuration that meets slack has a batch size larger than the number of invocations available for a given operation (using Equation~\ref{eq:objective}), it waits until more invocations arrive to assign a configuration.
It does so \emph{safely} --- only if there are enough upstream invocations and slack will not be violated.
Otherwise, it uses the best feasible configuration.

\myparagraph{Priority-based commit. }
Priority-based commit addresses the challenges of under-allotting slack and operations' affinity to heterogeneous hardware. 
First, to address the challenge of under-allotting slack, the Configurator prioritizes invoking a certain number of reference invocations for each operation, favoring deeper operations in the pipeline.
This ensures the feedback loop for all reference configurations is closed as fast as possible to minimize under-allotted slack.

Second, to compute an operation's affinity to heterogeneous hardware, \SystemName compares the benefits an operation invocation receives from running on a specific backend to other available backends. 
It computes the \emph{affinity} of an invocation's operation $op$ to hardware backend $\lambda$ using Equation \ref{eq:affinity}, where $X_{op,\lambda}$ is the subset of configurations for $op$ that run on $\lambda$ and $X_{op,\lambda}^{\mathsf{c}}$ is the complementary set (i.e., all other configurations for $op$). 

\begin{equation} \label{eq:affinity}
  affinity(op, \lambda) = \frac{\min_{\forall x \in X_{op,\lambda}^{\mathsf{c}}} \{obj(x, slack)\}}{\min_{\forall x \in X_{op,\lambda}} \{obj(x, slack)\}}
\end{equation}

Intuitively, Equation~\ref{eq:affinity} determines if a hardware backend provides more benefit (via Equation~\ref{eq:objective}) to an invocation than other backends.
\SystemName prioritizes invocations from operations with a higher affinity to a hardware backend $\lambda$ when committing them to each $CQ[\lambda]$.
This ensures each backend achieves its highest utility. 

\subsection{Handling stragglers and invocation failures} \label{sec:fault-tol}
During execution, operation invocations may straggle or fail to execute~\cite{nsdi13-dolly,socc14-wrangler,osdi04-mapreduce}.
The Scheduler keeps track of each invocation's execution time. 
If it exceeds a configurable time-out (discussed in Section~\ref{sec:implementation}) or the Scheduler receives an error, the Scheduler notifies the Manager to create a duplicate invocation.
This duplicated invocation is then passed to the Configurator to begin the slack allotment and configuration process anew.
The allotted slack will now be reduced, potentially resulting in a different configuration to still meet the pipeline latency target (evaluated in Section~\ref{sec:misprofiling}).

\section{Implementation} \label{sec:implementation}
We implemented \SystemName as an extension to gg in $\sim$4K lines of C++ code.
We modified gg's C++ and Python SDK to support complex pipelines and general knob configurations.
\SystemName supports operations from any framework or library; we implemented non-deep learning pipeline operations (e.g., blur and meanshift) using OpenCV~\cite{opencv_library} and FFmpeg~\cite{ffmpeg}, and deep learning pipeline operations with TensorFlow~\cite{tensorflow}.
\SystemName's source code will be available upon publication.

We implemented the online phase on top of gg's dispatcher and backend resource manager.
The online phase is single-threaded, but can scale out to multiple threads as needed.
For straggler mitigation, we set each invocation's time-out value to 1.5$\times$ the invocation's profiled latency.
Larger values wait too long to spawn a duplicate invocation, which may violate the pipeline latency target, while smaller values unnecessarily overload the speculation queues. 
For depth-first priority, we observed that 10 invocations of the reference configuration were sufficient to obtain enough feedback values to converge on a latency measurement.
Smaller values do not collect enough feedback values to prevent under-allotted slack, while larger values unnecessarily prioritize invocations with configurations that may not be efficient.

For the offline specification phase, we implemented the Operation-Profiler as a client to the online phase that collects and stores the profiled metadata into configuration specifications.
Configuration specifications are implemented as JSON files.
The Metadata Store is implemented in an object store (e.g., Google Cloud Storage).

We deployed \SystemName with serverless CPUs and serverless GPUs as compute backends.
For serverless CPUs, we provision and manage a cluster of CPUs similar to existing serverless offerings~\cite{atc18-peeking}.
Each invocation requests a specific number of cores (up to 4).
\SystemName also supports running on serverless computing services such as AWS Lambda~\cite{awslambda} or Google Cloud Functions~\cite{gcf}, where the invocation resources requested would be an amount of DRAM.

Since there exists no serverless GPU services or frameworks at the time of writing, we built our own implementation ($\sim$1K lines of C++ code) that we believe is representative of a future production offering~\cite{hotchips-gpu}.
Similar to CPU serverless offerings, an invocation requests an amount of GPU memory (in MB) per invocation.
Our serverless GPU scheduler then allocates a proportional amount of GPU threads using Nvidia MPS~\cite{mps}, allowing for multiple invocations to execute concurrently.
Invocations are executed on a first-come, first-served basis.
\SystemName is also compatible with GPUs that support concurrent job execution in hardware~\cite{a100}.

\section{Evaluation} \label{sec:evaluation}
\begin{table*}[t]
  \centering
  \small
  \resizebox{\linewidth}{!}{%
  \begin{tabular}{l|llll}
  \textbf{Pipeline} & \textbf{Description} & \textbf{Length (Form)} & \textbf{Operations (\# of total configurations)} & \textbf{Video input} \\ \hline
  AMBER Alert   & detect cars and people & 5 (branching) & \makecell{decode\textsuperscript{$\dagger$}, preprocess\textsuperscript{$\dagger$}, object detect., \\ face recog., car recog. (646)} & traffic camera~\cite{traffic}, 10 min, 1080p \\ \hline
  Face Blurring & detect indiv. face and blur from all frames & 5 (branching) & \makecell{decode\textsuperscript{$\dagger$}, preprocess\textsuperscript{$\dagger$}, face recog., \\ template match\textsuperscript{$\dagger$}, blur\textsuperscript{$\dagger$} (600)} & rally~\cite{bernie}, 10 min, 720p \\ \hline
  Denoising & detect indiv. face and denoise/segment & 5 (branching) & \makecell{decode\textsuperscript{$\dagger$}, preprocess\textsuperscript{$\dagger$}, face recog., \\ template match\textsuperscript{$\dagger$}, meanshift\textsuperscript{$\dagger$} (600)} & rally~\cite{bernie}, 10 min, 720p \\ \hline
  Toonify & apply cartoon effect to video & 4 (parallel) & \makecell{decode\textsuperscript{$\dagger$}, edge detect.\textsuperscript{$\dagger$}, bilateral filter\textsuperscript{$\dagger$}, \\ merge edge-filter\textsuperscript{$\dagger$}, encode\textsuperscript{$\dagger$} (989)} & tears of steel~\cite{tears-steel}, 10 min, 720p \\ \hline
  Synthetic & synthetic pipeline for sensitivity analysis & 7 (sequential) & \makecell{decode\textsuperscript{$\dagger$}, blur\textsuperscript{$\dagger$}, preprocess\textsuperscript{$\dagger$}, \\ face recog. (596)} & rally~\cite{bernie}, 10 min, 720p \\ \hline
  \end{tabular}
  } %
  \caption{\small Video pipelines used for evaluating \SystemName, their operations, and video inputs.
  $\dagger$ are non-deep learning pipeline operations.
  }
  \label{tab:pipeline_table}
  \vspace{-6mm}
\end{table*}

We answer the following questions:
(a) How does \SystemName compare to state-of-the-art systems (Scanner, Nexus, gg, and GrandSLAm)?
(b) How effective is \SystemName in meeting diverse latency targets?
(c) How does each technique employed by \SystemName, such as early speculation and late commit and safe delayed batching, contribute to its ability to meet the latency target?
(d) What is the impact of profiling errors and failures on \SystemName's ability to meet latency targets?
(e) What are the overheads of various decisions \SystemName makes? 

\myparagraph{Metrics. } Unless otherwise noted, we use pipeline processing latency and cost as metrics for success (similar to~\cite{eurosys19-grandslam,socc18-sprocket,socc20-inferline}).
For each experiment, we report the mean of three runs.

\myparagraph{Experimental setup. }
We deployed \SystemName on Google Cloud Platform (GCP)~\cite{gcp}.
The \SystemName runtime ran on a \texttt{n1-standard-8} instance (8 vCPUs, 30 GB of DRAM).
We used the following setup unless otherwise noted.
For the serverless CPU backends, we used 10 \texttt{n1-standard-64} (64 vCPUs, 240GB of DRAM).
For the serverless GPU backends, we used 2 \texttt{custom-12-46080} (1 V100 GPU, 12 vCPUs, 45 GB of DRAM).
All instances feature Intel Xeon Platinum E5-2620 CPUs operating at 2.20GHz, Ubuntu 16.04 with 5.3.0 kernel, and up to 32 Gbps networking speed.

\myparagraph{Baseline systems. }
We compared \SystemName with three sets of systems: (a) cluster systems (Scanner and Nexus), (b) serverless systems (gg), and (c) target-aware systems (GrandSLAm).
Scanner is used by Facebook for processing 360$^\circ$ videos~\cite{scanner-website}.
Nexus accelerates deep learning-based video analysis on GPUs.
gg is a general purpose serverless framework.
GrandSLAm estimates slack to meet pipeline latency targets for sequential, DNN-only pipelines.
We evaluated two common Scanner setups: one in which a user only provisions a cluster with CPUs (\texttt{sc-cpu}), and one in which, similar to Nexus, a user runs all operations on a GPU (\texttt{sc-gpu}).
For gg, we also compared against a version augmented with \SystemName's branching support (\texttt{gg-branch}). 
\texttt{sc-cpu}, \texttt{gg}, and \texttt{gg-branch} do not support heterogeneous accelerators, while \texttt{Nexus} and \texttt{sc-gpu} require GPU VMs.
Since GrandSLAm does not natively support non-sequential pipelines, and does not account for input-dependent execution flow, we implement it with \SystemName by disabling early speculation and late commit, feedback, and depth-first priority.
However, GrandSLAm still has access to \SystemName's branching support, safe delayed batching, priority-based commit, and dynamic resource allocation across heterogeneous backends (\texttt{GrandSLAm++}).
To equalize compute resources provided to all systems, we observe that \texttt{custom-12-46080} and \texttt{n1-standard-64} VMs are effectively priced the same on GCP (a difference of 1\% at the time of writing).
(This price-equivalency is also true for equivalent instances on AWS.)
Thus, we provisioned \texttt{sc-cpu}, \texttt{gg}, \texttt{gg-branch}, and \texttt{GrandSLAm++} with 12 \texttt{n1-standard-64}, and \texttt{Nexus} and \texttt{sc-gpu} with 12 \texttt{custom-12-46080}.

\myparagraph{Resource requests and cost model. }
For \SystemName, gg, and \texttt{GrandSLAm++}, each invocation requests a set amount of resources (GPU memory or CPU cores) as is done in commercial serverless offerings.
The respective backend then provisions the invocation with the requested resources, and charges a price based on the amount of requested resources and invocation latency.
We calculate the price (in \$/(resource-second)) by dividing the cost per second charged by GCP by the VM's total resources.
For example, the price of a V100 GPU invocation is calculated by dividing the price of \texttt{custom-12-46080} by $16GB$.
Since Scanner and Nexus are cluster-based frameworks, we compute cost using the time to rent the cluster for the duration of the execution; we do not include the cost of starting and maintaining a warm cluster.

\myparagraph{Video pipeline. }
Table~\ref{tab:pipeline_table} shows the pipelines, operations, and videos that we used.
For branching pipelines, only invocations satisfying the branching condition are executed.
For AMBER Alert, only frames with faces and cars execute their respective recognition paths. %
For Face Blurring and Denoising, frames with faces proceed to a template match operation where the frame is compared against a pre-determined face.
If a match is found, the face in the frame is then blurred, or denoised using meanshift.
The Toonify pipeline executes the bilateral filtering and edge operations in parallel before merging and encoding the frames.
Finally, the synthetic pipeline is a chain of 5 image blurring operations followed by a face recognition operation.
The face recognition operation is the most compute-intensive operation of this pipeline, which allows us to evaluate \SystemName's ability to meet diverse pipeline latency targets, even when configurations were mis-profiled (Section \ref{sec:misprofiling}).
Since \texttt{sc-cpu}, \texttt{sc-gpu}, and \texttt{gg} do not support branches, they execute the three branching pipelines as parallel ones (i.e., both branches are executed).

\begin{figure}[t]
  \centering
  \includegraphics[width=.95\columnwidth]{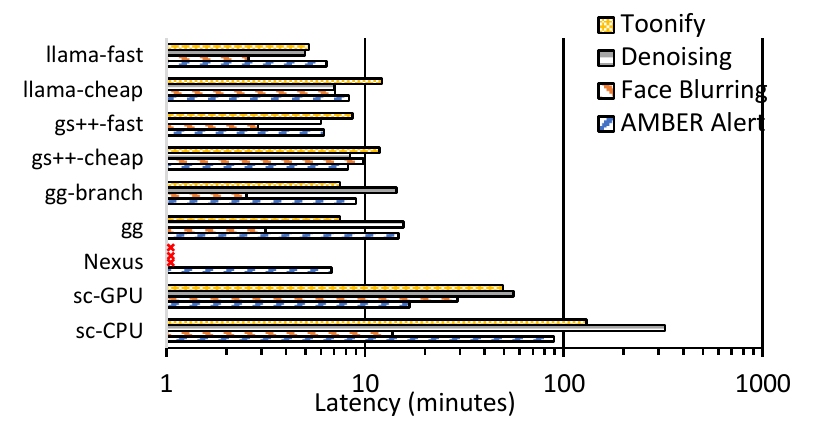} 
  \caption{\small Latency of baselines to execute each pipeline.
    Nexus only supports the AMBER Alert pipeline (unsupported pipelines are denoted by $\times$).
    \SystemName's fastest execution is faster than all baselines.}
  \label{fig:latency-graph}
\end{figure}

\begin{figure}[t]
  \centering
  \includegraphics[width=.95\columnwidth]{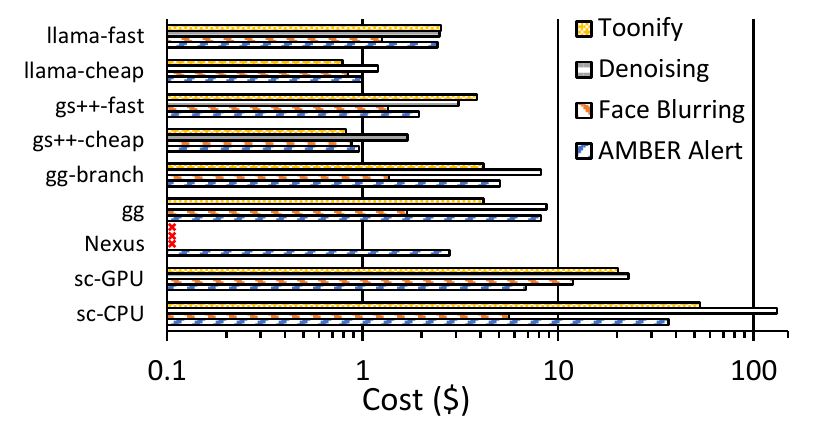} 
  \caption{\small Cost incurred by baselines for each pipeline.
    Nexus only supports the AMBER Alert pipeline (unsupported pipelines denoted by $\times$).
    \SystemName's cheapest execution is cheaper than all baselines.}
  \label{fig:cost-graph}
\end{figure}

\subsection{Comparing \SystemName to existing systems} \label{sec:sota}
We first show how \SystemName's ability to dynamically reconfigure operation invocations enables it to outperform existing systems, both in terms of latency and cost.

\myparagraph{Experimental setup. }
For \texttt{Nexus}, we set the pipeline latency target to be 2 seconds per frame, which we found to be the strictest latency that does not drop any requests~\cite{sosp19-nexus}.
\texttt{Nexus} then automatically configures the batch size and number of instances for each model.
For \texttt{sc-cpu} and \texttt{sc-gpu}, we swept each operation's batch size from 1 to 64 (by powers of 2) and set each value based on the lowest pipeline execution latency (reported in Figure~\ref{fig:latency-graph}).
For \texttt{gg} and \texttt{gg-branch}, we set each invocation's configuration based on the lowest, most cost-effective CPU latency reported by the Operation-Profiler.
We configured \SystemName and \texttt{GrandSLAm++} with two pipeline latency targets: an unachievable low target that forced both to minimize pipeline execution latency at the expense of cost: \texttt{llama-fast} and \texttt{GrandSLAm++-fast}, and an overly-loose target that allowed \SystemName to minimize the overall cost: \texttt{llama-cheap} and \texttt{GrandSLAm++-cheap}.

\myparagraph{Results and discussion. }
Figures \ref{fig:latency-graph} and \ref{fig:cost-graph} show the processing latency and total cost, respectively, to execute each of the four non-synthetic pipelines.
\SystemName achieves lower latency, higher throughput, and lower cost than existing systems.

Even when the cost of starting and maintaining a warm cluster are not considered,
\SystemName is faster (up to 65$\times$ and 28$\times$ on average) and cheaper (up to 110$\times$ and 55$\times$ on average) than \texttt{sc-cpu}.
Compared to \texttt{sc-gpu}, \SystemName is up to 11$\times$ faster (6$\times$ on average) and up to 27$\times$ cheaper (18$\times$ on average).
Scanner cannot dynamically adjust and right-size invocation configurations, and thus cannot address performance degradation caused by resource contention for compute-intensive operations (e.g., deep learning inference and meanshift) or memory-intensive operations (e.g., bilateral filtering).

Next, since \texttt{Nexus} focuses on inference-serving pipelines, we are only able to run the AMBER Alert pipeline (other pipelines denoted by $\times$ in Figures~\ref{fig:latency-graph} and~\ref{fig:cost-graph}).
While we provide \texttt{Nexus} with 12 GPUs, \texttt{Nexus}'s bin-packing algorithm~\cite{sosp19-nexus} utilizes only 8; thus we report cost for 8 GPUs.
By dynamically choosing CPU versus GPU configurations, \SystemName achieves 1.3$\times$ speedup and 2.8$\times$ lower cost compared to \texttt{Nexus}.

Compared to \texttt{gg}, \SystemName is up to 3.1$\times$ faster (2.2$\times$ on average) and up to 8.2$\times$ cheaper (5.7$\times$ on average).
Compared to \texttt{gg-branch}, \SystemName is up to 2.9$\times$ faster (1.8$\times$ on average) and up to 6.8$\times$ cheaper (4.7$\times$ on average).
While \texttt{gg-branch} can reason about conditional flow, it cannot make dynamic invocation configuration decisions or adjust to resource volatility, resulting in a higher latency and cost compared to \SystemName.

Finally, \SystemName is up to 1.7$\times$ faster (1.2$\times$ on average) than \texttt{GrandSLAm++-fast} and 1.4$\times$ cheaper (1.1$\times$ on average) than \texttt{GrandSLAm++-cheap}.
Since \texttt{GrandSLAm++} allots slack and selects configurations based on profiled values, it cannot dynamically adjust to nondeterminism, which can result in slower performance or higher cost (e.g., Denoising).

By making dynamic invocation configurations, \SystemName is able to determine how well operations perform across heterogeneous backends and right-size resources depending on the pipeline latency target.

\myparagraph{General applicability. }
While \SystemName was designed to address the challenges of running video analytics and processing pipelines (Section~\ref{sec:motivation}), its operation configuration specification (Section~\ref{sec:offline}) supports arbitrary operation- and hardware-specific configuration knobs.
To demonstrate this, we built a four-stage natural language processing pipeline for applications like therapy session analysis for at-risk youth~\cite{affcomp2019-behavior}.
The pipeline has six models (language identification, two language translation, sentiment analysis, text generation, and summarization) and features branching, parallel, and sequential patterns.
For a 256-line transcript, \texttt{llama-fast} takes 275s (\$1.22) while \texttt{llama-cheap} takes 573s (\$0.44).
Thus, \SystemName can be used for meeting latency targets for general domains.

\begin{figure}[t]
  \centering
  \includegraphics[width=.95\columnwidth]{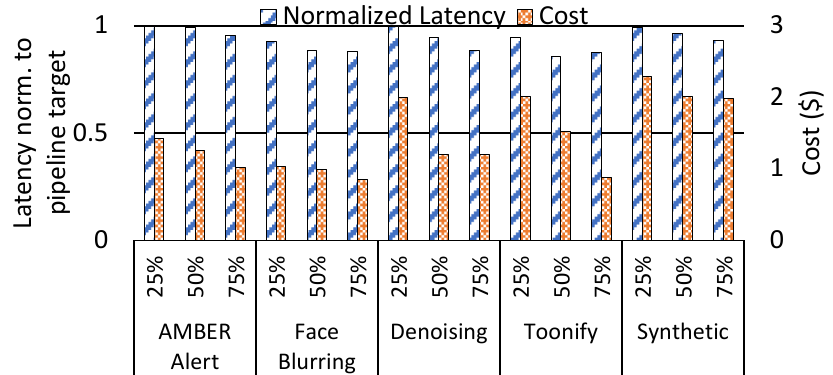}
  \caption{\small Evaluating \SystemName given varied latency targets. 
  50\%: mean of the measured latencies of \texttt{llama-fast} and \texttt{llama-cheap}, 25\%: mean of \texttt{llama-fast} and 50\%, and 75\%: mean of \texttt{llama-cheap} and 50\%.
  The execution latency is normalized to the pipeline target ($\leq$1 means target was met).
  Cost is in dollars. %
  \SystemName meets all latency targets and reduces overall cost for less stringent targets.}
  \label{fig:sweep} 
  \vspace{-6mm}
\end{figure}

\begin{table}[t]
  \centering
  \small
  \resizebox{\linewidth}{!}{%
  \begin{tabular}{llllll}
  \textbf{Pipeline} & \textbf{\# configs. used} & \textbf{\% invoc. that met slack} \\ \hline
  AMBER Alert & 27 $\pm$ 8 & 92\% $\pm$ 6\% \\
  Face Blurring & 29 $\pm$ 3 & 93\% $\pm$ 1\% \\
  Denoising & 40 $\pm$ 4 & 99\% $\pm$ 0\% \\
  Toonify & 30 $\pm$ 5 & 97\% $\pm$ 3\% \\
  Synthetic & 50 $\pm$ 16 & 88\% $\pm$ 3\% \\ \hline
  \end{tabular}
  } %
  \caption{\small Mean and standard deviation of number of configurations used and percent of invocations that met their allotted slack.
  \SystemName accurately allots and meets almost all slack by selecting a variety of different configurations per pipeline.
  }
  \label{tab:insights}
  \vspace{-8mm}
\end{table}

\subsection{Can \SystemName trade off latency for cost?} \label{sec:lattargs}
We now show \SystemName can also meet latency targets that lie between \texttt{llama-fast} and \texttt{llama-cheap}.

\myparagraph{Experimental setup. }
For each pipeline, we provide three latency targets to \SystemName that lie between the times required to execute the pipeline using \texttt{llama-fast} and \texttt{llama-cheap}.
The 50\% latency target is the mean latency between the latencies achieved by \texttt{llama-fast} and \texttt{llama-cheap}.
The 25\% latency target (the most stringent of the three) is the mean latency between \texttt{llama-fast} and the 50\% latency target.
Finally, the 75\% latency target (the least stringent of the three) is the mean latency between \texttt{llama-cheap} and the 50\% latency target.
For example, \texttt{llama-fast} executed Face Blurring in 155 seconds, and \texttt{llama-cheap} executed it in 423 seconds; the 25\%, 50\%, and 75\% latency targets are 225, 290, and 380 seconds respectively.

\myparagraph{Results and discussion. }
Figure~\ref{fig:sweep} shows the observed execution latency, normalized to each of the aforementioned pipeline latency targets ($\leq$1 means that the latency target was met), as well as the raw cost values for each pipeline execution.
\SystemName not only meets all latency targets, but also dynamically adjusts its configuration decisions to choose cost-efficient configurations as the latency target became less stringent.
For the Denoising and Synthetic pipelines, the cost stays the same for the 50\% and 75\% targets.
This is due to \SystemName selecting similar invocation configurations during both runs, since it determined them to be the most cost-efficient configurations for both latency targets.

Table~\ref{tab:insights} shows a breakdown of how many configurations were used to meet the 50\% pipeline latency target, and what percent of invocations met the slack.
We note that (a) \SystemName meets the slack for 94\% of invocations on average across all pipelines, with the lowest being the Synthetic pipeline since it is the longest, and (b) the number of configurations used varies per pipeline.
Thus, \SystemName's slack allottment and configuration selection algorithms (Section~\ref{sec:online}) are effective in meeting pipeline latency targets while minimizing cost.

\begin{figure}[t]
  \centering
  \includegraphics[width=.95\columnwidth]{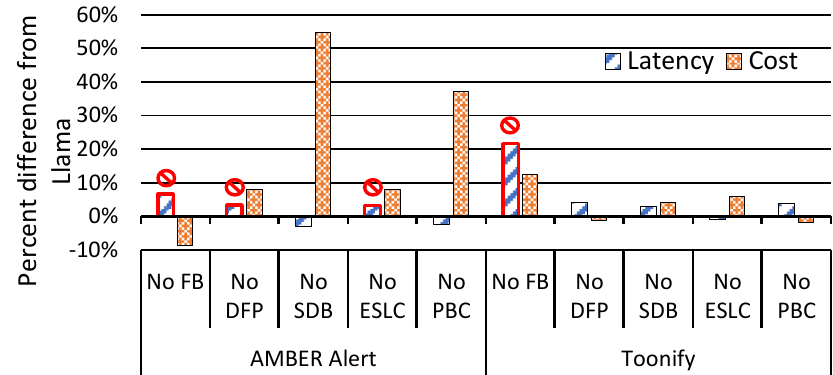} 
  \caption{\small Impact of turning \SystemName's techniques off on the AMBER Alert and Toonify pipelines. %
  Red borders and circled slashes indicate the pipeline latency target was violated.
  \textbf{FB} is feedback, \textbf{DFP} is depth-first priority, \textbf{SDB} is safe-delayed batching, \textbf{ESLC} is early speculation and late commit, and \textbf{PBC} is priority-based commit.}
  \label{fig:ablation} 
  \vspace{-6mm} 
\end{figure}

\subsection{Ablation study of \SystemName's techniques} \label{sec:ablation}
We now show how each technique of \SystemName contributes to its ability to efficiently meet pipeline latency targets.

\myparagraph{Experimental setup. }
We performed an ablation study with two distinct pipelines: Amber Alert and Toonify. 
Following is the list of techniques employed by \SystemName: feedback loop (FB, Section~\ref{sec:onlinepreview}), depth-first priority (DFP, Section~\ref{sec:inter-operation}), safe delayed batching (SDB, Section~\ref{sec:inter-operation}), early speculation and late commit (ESLC, Section~\ref{sec:queueing}), and priority-based commit (PBC, Section~\ref{sec:inter-operation}).
Note that priority-based commit includes both depth-first priority and hardware affinity.
For each run, we turn off a single technique and record the pipeline execution latency and cost.
For each pipeline, we use its 50\% pipeline latency target specified in Section \ref{sec:lattargs}.

\myparagraph{Results and discussion. }
Figure~\ref{fig:ablation} shows the results of our ablation study (red borders and circled slashes indicate the latency target was violated).
For the AMBER Alert pipeline, disabling feedback, depth-first priority, or early speculation and late commit results in latency target violations.
All three techniques allow \SystemName to accurately measure and adapt to performance volatility caused by input-dependent execution flow (branching operations) and resource contention.
For example, disabling feedback causes \SystemName to miss the latency target because resource contention resulted in invocations taking longer than their profiled values. 
With feedback enabled, \SystemName is able to detect this and choose configurations with higher throughput at a small expense of cost-efficiency.
On the other hand, disabling safe delayed batching or priority-based commit causes \SystemName to not use large batches for deep learning inference invocations on GPUs, resulting in reduced cost-efficiency.

For the Toonify pipeline, disabling feedback also causes a latency target violation similar to the AMBER Alert pipeline.
Disabling either safe delayed batching or early speculation and late commit results in \SystemName choosing less cost-efficient configurations.
On the other hand, disabling depth-first priority and priority-based commit results in more cost-efficient configurations without violating the latency target.
This is because these techniques led to \SystemName choosing configurations that are more throughput-intensive than necessary for merge edge\-/filter operation invocations in an effort to meet the pipeline latency target.
However, as noted for the AMBER Alert pipeline and evaluated in Section \ref{sec:misprofiling}, both depth-first priority and priority-based commit are important for \SystemName's robustness in right-sizing resources and meeting latency targets despite profiling errors.

\begin{table}[t]
  \centering
  \small
  \resizebox{\linewidth}{!}{%
  \begin{tabular}{lll}
  \textbf{Pipeline (target)} & \textbf{\SystemName} & \textbf{\SystemName w/o FB \& DFP} \\ \hline
  Denoising (350s)  & (348s, \$1.20) & (369s, \$1.64) \\
  Synthetic (520s) & (520s, \$2.31) & (487s, \$3.14) \\ \hline
  \end{tabular}
  } %
  \caption{\small Impact of profiling errors. Latency and cost for the Denoising and Synthetic pipelines when profiled values are inaccurate (set to 50\% of their measured latencies).
  \textbf{FB} is feedback and \textbf{DFP} is depth-first priority.
  Without these techniques, \SystemName cannot meet the latency target, or uses configurations that are not cost-effective.
  }
  \label{tab:mis-profiling} 
  \vspace{-10mm}
\end{table}

\subsection{Meeting targets despite profiling errors \& failures} \label{sec:misprofiling}
We now show that \SystemName can meet targets despite profiling errors and invocation failures.

\myparagraph{Experimental setup. }
To evaluate ``mis-profiling'', all operation profiled latencies are set to 50\% of their values.
Separately, to evaluate \SystemName's resiliency to failures, we forced $3\%$ of invocations to fail ($2,114$ and $3,617$ failures for the Denoising and Synthetic pipeline, respectively).
For both experiments, we used the Denoising and Synthetic pipelines because they represent worst-case scenarios: an expensive operation at the end of the pipeline with an under-estimated latency. 
In addition, the Synthetic pipeline is the longest pipeline, which further exacerbates profiling errors: \SystemName will under-allot slack to the last operation unless techniques are used to mitigate mis-profiling.
We use the respective 50\% pipeline latency target from Section \ref{sec:lattargs} for each pipeline.

\myparagraph{Results and discussion. }
Table \ref{tab:mis-profiling} shows the impact of profiling error on latency and cost with (a) all of \SystemName's techniques, and (b) both feedback and depth-first priority turned off (the two techniques \SystemName relies on to adjust for inaccurate profiling).
For the Denoising pipeline, disabling feedback and depth-first priority causes \SystemName to under-allot slack to the last meanshift operation.
This results in a missed pipeline latency target because \SystemName could not adjust to the profiling errors until late in the pipeline execution.
For the Synthetic pipeline, when both techniques were off, \SystemName meets the latency target but at a 35\% higher cost.
This is because the 50\% lower-than-profiled latencies cause \SystemName's objective function (Equation~\ref{eq:objective}) to incorrectly calculate that the CPU, not the GPU, is most cost-efficient for the meanshift operation.
Since the CPU configuration was actually 50\% slower and less cost-effective than a GPU, the cost ends up being higher than if a GPU would have been used.

When evaluating invocation failures, both pipelines were able to meet the specified latency target despite the high failure rate using the techniques described in Section~\ref{sec:fault-tol}.

These results demonstrate depth-first priority and feedback are necessary to resolve profiling discrepancies early on during execution, and that \SystemName is robust to failures.

\begin{table}[t]
  \centering
  \footnotesize
  \begin{tabular}{lll}
    \textbf{Phase} & \textbf{Action} & \textbf{Latency (\% of exec.)}           \\ \hline
    Specification & Profiling            & 257 $\pm$ 155 s      \\
                  & Path decomposition   & 1.74 s               \\ \hline
    Online        & Speculate            & 0.005 $\pm$ 0.005 ms (0.08\%) \\
                  & Commit               & 0.186 $\pm$ 0.813 ms (3.1\%) \\
                  & Invoke               & 0.151 $\pm$ 0.078 ms (2.5\%) \\
                  & Finalize             & 0.141 $\pm$ 1.209 ms (2.4\%) \\ \hline
  \end{tabular}
  \caption{\small \SystemName's decision overheads. Mean and standard deviation latencies of invocations for the AMBER Alert pipeline. %
    Latencies are per-invocation for online actions, per-operation for profiling, and per-pipeline for path decomposition.
    For each online action, we show the percent of the execution time spent on the action across all operation invocations (73K).
  }
  \label{tab:overheads}
  \vspace{-4mm}
\end{table}

\subsection{Overheads of decisions \SystemName makes} \label{sec:scale} 
Finally, we evaluate \SystemName's overheads and its ability to scale across backends.

Table \ref{tab:overheads} shows the overhead for these decisions when specifying and running the AMBER Alert pipeline with the 50\% intermediate latency target from Section~\ref{sec:lattargs}; all other pipelines have similar overheads.
For the specification phase, profiling each operation takes an average of 257 seconds, and only needs to be performed the first time an operation is added to the Metadata Store.
The decomposition step, which is performed once per pipeline, takes only 1.7 seconds.

During the online phase, \SystemName only spends 483 micro-seconds, on average, to process (i.e., speculate, commit, invoke, and finalize) an invocation, allowing \SystemName to process over 2000 invocations per second.
Calculating a slack and determining a configuration is efficient, as speculate only requires 5 micro-seconds.
Most time is spent evaluating priority between operations during commit, connecting and sending invocations to backends during invoke, and updating global state once invocations completed during finalize.

Low overheads also allows \SystemName to improve execution latency as the number of resources or maximum concurrency increases.
Compared to \texttt{llama-fast} for the AMBER Alert pipeline run on 10 CPU and 2 GPU instances (Section~\ref{sec:sota}), having 6 CPU and 1 GPU instances is 46\% slower, while having 15 CPU and 3 GPU instances is 25\% faster.

\section{Related Work} \label{sec:related-work}
\myparagraph{Video and general-purpose analytics frameworks. }
In Section~\ref{sec:motivation}, we describe the limitations of several existing video analytics and processing frameworks~\cite{sosp19-nexus,osdi18-focus,siggraph18-scanner,nsdi17-videostorm,socc20-inferline,atc19-gg,socc18-sprocket}.
Other cluster-based and serverless systems for both domain-specific and general-purpose applications~\cite{hotcloud10-spark, eurosys12-jockey, atc18-videochef, eurosys07-dryad, nsdi11-ciel, sigmod10-pregel, socc17-pywren} either do not support independent-dependent execution flow, require extensive per-pipeline profiling, or require users to configure and right-size resources.
\SystemName meets diverse pipeline latency targets across complex (and possibly dynamic) video pipelines using heterogeneous serverless backends.

\myparagraph{Dataflow optimizations and scheduling techniques. }
GrandSLAm~\cite{eurosys19-grandslam} and Fifer~\cite{middleware20-fifer} use slack to statically determine the batch size for sequential microservice graphs.
Delayed batching is used by Clipper~\cite{nsdi17-clipper} to increase efficiency of inference queries, but must be statically set by users.
Late binding is used by schedulers~\cite{nsdi20-sol, sosp13-sparrow, eurosys18-bindschaedler, osdi18-qoop, osdi14-venkataraman} to maximize the flexibility of the scheduling decision and knowledge of system state.
However, these systems do not consider the need to configure operations for meeting pipeline latency targets.
TetriSched~\cite{eurosys16-tetrisched} uses a scheduler that prevents tasks from being sent to a sub-optimal set of resources due to resources being held by earlier jobs, but only supports per-operation targets, not an end-to-end pipeline latency target.
Early speculation and late commit, and priority-based commit allow \SystemName to compute slack and make configuration decisions for arbitrarily complex pipelines to meet overall pipeline latency targets.
Musketeer~\cite{eurosys15-musketeer} and Dandelion~\cite{sosp13-dandelion} optimize dataflow DAGs for execution on a broad range of execution engines or hardware platforms. 
These optimizations are compatible with \SystemName, and can be used to expand the backends and hardware platforms \SystemName supports.

\myparagraph{Cost-based query optimization. }
Several works have explored cost-based query optimization for relational databases~\cite{sigmod02-rbqo,vldb19-lbce,vldb12-dbtoaster,vldb18-listopt,vldb19-ibtune,vldb11-mrp}, including for queries whose optimal plan is input-dependent~\cite{vldb20-tempura}.
\SystemName is compatible with these frameworks, and can leverage their optimizations as an extension to how configurations are selected.

\myparagraph{Auto-tuning configurations. }
CherryPick~\cite{nsdi17-cherrypick} and Ernest~\cite{nsdi16-ernest} present a performance prediction framework for recurring data analytics jobs; however, these systems require tens of executions of a job to set the configuration parameters. 
PARIS~\cite{paris} focuses on VM-size selection; OptimusCloud~\cite{atc20-optimuscloud} and Selecta~\cite{atc18-selecta} are domain-specific VM configuration systems for databases and storage technologies, respectively. 
\SystemName dynamically configures general video operations to meet diverse latency targets, and only requires one-time per-operation profiling.

\vspace{-4mm}
\section{Conclusion} \label{sec:conclusion} 
We presented \SystemName, a heterogeneous and serverless video analytics and processing framework that executes general video pipelines, meeting user-specified performance targets at minimal cost.
By dynamically configuring individual operation invocations, \SystemName efficiently traverses large configuration spaces, adapts to input-dependent execution flow, and dynamically allocates resources across heterogeneous serverless backends.
\SystemName makes per-operation invocation decisions by first calculating invocation slack, then leveraging techniques such as safe delayed batching, priority-based commit, and early speculation and late commit to efficiently and accurately select configurations that meet the slack. 
\SystemName achieves an average improvement of 7.8$\times$ for latency and 16$\times$ for cost compared to state-of-the-art systems.

\bibliographystyle{abbrv} %
\bibliography{paper}

\end{document}